\documentclass[a4paper,11pt]{article}

\usepackage[utf8]{inputenc} 
\usepackage[a4paper]{geometry}
\usepackage{xarticle}
\usepackage{makecell}

\usepackage[T1]{fontenc}
\usepackage{ccfonts}
\usepackage{listings}
\usepackage{paralist}
\usepackage[parfill]{parskip}
\usepackage{eurosym}
\usepackage{hyperref}
\usepackage{listings}
\usepackage{paralist}
\usepackage{graphicx}
\usepackage{color}
\usepackage{threeparttable}
\usepackage{pdflscape}

\definecolor{dkgreen}{rgb}{0,0.6,0}
\definecolor{gray}{rgb}{0.5,0.5,0.5}
\definecolor{mauve}{rgb}{0.58,0,0.82}
\definecolor{lightgray}{gray}{0.85}

\lstdefinelanguage{metadata}{
	sensitive=true,
	classoffset=1,
	keywords=[1]{name,version,setup,requiredMemory,priority,
	commands,command,arg,arguments,type,required,outputs,
	description,input_name,value,uri,outputType,isRequired,argument_name, builder},
	 keywordstyle=\color{blue},
	   classoffset=2,
 morekeywords={outputDir,true, directory_dependent},
  keywordstyle=\color{mauve},
morecomment=[s]{/*}{*/},
morecomment=[l][keywordstyle4]{\#}
}

\lstdefinelanguage{ngspipes}{
	sensitive=true,
	classoffset=1,
	keywords=[1]{Pipeline,tool,command,argument,chain},
	 keywordstyle=\color{blue},
	   classoffset=2,
 morekeywords={Github,DockerConfig},
  keywordstyle=\color{mauve},
morecomment=[s]{/*}{*/},
morecomment=[l][keywordstyle4]{\#}
}

\newcommand{\footremember}[2]{%
   \footnote{#2}
    \newcounter{#1}
    \setcounter{#1}{\value{footnote}}%
}
\newcommand{\footrecall}[1]{%
    \footnotemark[\value{#1}]%
}
\def\paperid{\small\hfill}
\SquashBibFurther
\begin{document}
\title{Beyond NGS data sharing and towards open science}
\author{%
    Bruno Dantas\footremember{a}{Instituto Superior de Engenharia de Lisboa, Instituto Polit\'{e}cnico de Lisboa}
    \and Calmenelias Fleitas\footrecall{a}%
    \and Alexandre P Francisco\footremember{b}{Instituto Superior T\'{e}cnico, Universidade de Lisboa}\footremember{c}{Instituto de Engenharia de Sistemas e Computadores, I\&D}
    \and Jos\'{e} Sim\~{a}o\footrecall{a} \footrecall{c}
    \and C\'{a}tia Vaz\footrecall{a} \footrecall{c} %
}
\date{}

\maketitle

\begin{abstract} 
Biosciences have been revolutionized by next generation sequencing (NGS) technologies in last years, leading to new perspectives in medical, industrial and environmental applications.
And although our motivation comes from biosciences, the following is true for many areas of science: published results are usually hard to reproduce either because data is not available or tools are not readily available, which delays the adoption of new methodologies and hinders innovation.
Our focus is on tool readiness and pipelines availability.
Even though most tools are freely available, pipelines for data analysis are in general barely described and their configuration is far from trivial, with many parameters to be tuned.

In this paper we discuss how to effectively build and use pipelines, relying on state of the art computing technologies to execute them without users need to configure, install and manage tools, servers and complex workflow management systems.
We perform an in depth comparative analysis of state of the art frameworks and systems.
The NGSPipes framework is proposed showing that we can have public pipelines ready to process and analyse experimental data, produced for instance by high-throughput technologies, but without relying on centralized servers or Web services.

The NGSPipes framework and underlying architecture provides a major step towards open science and true collaboration in what concerns tools and pipelines among computational biology researchers and practitioners.
We show that it is possible to execute data analysis pipelines in a decentralized and platform independent way.
Approaches like the one proposed are crucial for archiving and reusing data analysis pipelines at medium/long-term.
NGSPipes framework is freely available at \href{http://ngspipes.github.io/}{http://ngspipes.github.io/}.
\end{abstract} 

\section*{Background}
Nowadays most scientific experiments that employ next-generation sequencing (NGS) rely on running and refining a series of intertwined computational analysis and visualization tasks on large amounts of data.
These so called analysis pipelines, or more generally workflows, start with voluminous raw sequences and can end with detailed structural, functional, and evolutionary results.
Pipelines involve the use of multiple software tools and data resources in a staged fashion, with the output of one tool being passed as input to the next one.
A personalized medicine pipeline based on NGS technology can start for instance with short DNA sequences (reads) of an individual human genome and end with a diagnostic and prognostic report~\cite{Koboldt2010,Voelkerding2008}.
It can even end with a treatment plan if clinical data are available.
This kind of pipelines depends on the use of multiple software tools to assess the quality of reads, map them to a reference human genome, identify sequence variations, query databases for the sake of associating variations to diseases, and check for novel variants.
All these tools must be parametrized, a task that is in general far from trivial and that may depend on costly research experiments.

The data and analysis protocols made available in database records and along publications are also almost never sufficient to reproduce analyses or assess results~\cite{Goodman2014}. 
%
%
At the same time, the amount of data generated in scientific experiments is outpacing the computational and storage capabilities available in most research labs.
This is especially true for life sciences, where new technologies increased the sequencing throughput from kilobytes to terabytes per day.
Experiments that employ NGS lead in general to challenges in reproducibility due to a lack of standards, exceedingly large dataset sizes, and increasingly complex computational tools.
The usage of multiple data sources and computational tools in these studies further complicate reproducibility.

To simplify the design and execution of biomedical workflows by end users, especially those that use multiple software tools and data resources, a number of scientific workflow systems have been developed over the past decade.
Scientific workflows correspond to series of structured activities and computations that arise in scientific problem solving.
Workflows are however more general than data analysis pipelines, allowing user interactions along their execution and possibly lasting for long periods of time.
Both involve the invocation of a number and variety of analysis tools.  
And one of the main aims of these systems is to allow the integration of both tools, services and data sources, without requiring programming experience.
The most used, and maintained systems and frameworks are listed and compared in Table~\ref{tablecomparison}.
The comparison provided, and comparison criteria, are an extension of the recent comparison by Leipzig~\cite{leipzig2016review}.
The most relevant criteria for our work are the containerization of tools and pipeline sharing.
Most of systems depend on installing tools before running pipelines.
And although pipelines can be shared among users, one must rely on some kind of virtual research environment offered through a centralized service. We are looking for a decentralized and self-contained approach. 

Workflows and pipelines are usually abstracted in existing systems as directed graphs, where nodes represent tasks to be executed and edges represent either the data flow or execution dependencies between different tasks.
Pipelines for data analysis are in general less complex and can be abstracted as directed acyclic graphs (DAG),
The system execution engine maps then the nodes in the graph to real data and software components, and is responsible for executing the software components in order, either locally on the user machine or remotely, for instance using cloud services.
Some of these scientific workflow systems may use high performance computing facilities, if available, for processing large volumes of data concurrently.
But most of these scientific workflow systems cannot be easily installed and configured, are available as centralized services, and are most of the times only available to users with access to some kind of specialized IT support.

\newgeometry{left=0.5cm,bottom=0cm,right=0cm,top=0cm}
\begin{landscape}
\thispagestyle{empty}
\begin{table}[!t]
    \centering
\caption{Comparison of tools and frameworks.\label{tablecomparison}}
{
\begin{threeparttable}
\begin{tabular}{|c|c|c|c|c|c|c|c|c|c|} 
\hline
& \multicolumn{1}{c|}{{\bf \thead{Tool\\dependency\tnote{a}} }}  & \multicolumn{1}{c|}{{\bf \thead{Execution\\order\tnote{b}}}} & \multicolumn{1}{c|}{{\bf Paradigm\tnote{c}} } & \multicolumn{1}{c|}{{\bf Workbench\tnote{d}}}  & \multicolumn{1}{c|}{{\bf Execution\tnote{e}}}  & \multicolumn{1}{c|}{{\bf \thead{Nested\\pipelines\tnote{f}}}}  & \multicolumn{1}{c|}{{\bf \thead{Pipeline\\sharing\tnote{g}}}} & \multicolumn{1}{c|}{{\bf Parallelization\tnote{h}} } & \multicolumn{1}{c|}{{\bf \thead{Containerization\\of tools\tnote{i}}} }\\  \hline
\bf{Arvados\cite{arvados}}       & explicit   & implicit  & configuration& script/GUI & CLI/GUI & yes & partial & multi-core/task & no  \\ \hline
\bf{BigDataScript\cite{cingolani2015bigdatascript}} & implicit  & explicit  & convention & script & CLI & no & partial & multi-core/task & no  \\ \hline
\bf{bpipe\cite{sadedin2012bpipe}}       & explicit  & explicit  & convention & script & CLI & no & no & multi-core & no  \\ \hline
\bf{Conveyor\cite{Linke2011}}       & explicit  & implicit  & configuration  &  GUI  & GUI & no & no  & multi-core/task & no  \\ \hline
\bf{DiscoveryNet\cite{Rowe2003}}       & explicit  & explicit  & configuration & GUI  & GUI & yes & yes & multi-core/task & no \\ \hline
\bf{Galaxy\cite{Giardine2005}}       & explicit  & explicit  & configuration & GUI  & GUI & no & partial & multi-core; data partition & no \\ \hline
\bf{GenePattern\cite{Reich2006}}       & explicit  & implicit  & configuration  & GUI  & GUI & no & partial & multi-core/task;data partition & no  \\ \hline
\bf{Kepler\cite{Ludascher2006}}       & explicit  & explicit  & configuration  & GUI  & GUI & yes & partial & multi-core/task & no  \\ \hline
\bf{Luigi\cite{luigi}}       & implicit  & implicit  & class-based & script & CLI/GUI & no & no & multi-core/task & no \\ \hline
\bf{NGSpipes}       & explicit  & implicit  & configuration & script/GUI   & CLI/GUI & no & yes & multi-core & yes  \\ \hline
\bf{NextFlow\cite{nextflow}} & implicit  & implicit & convention & script  & CLI  & no & partial & multi-core/task & yes \\ \hline
\bf{Pegasus\cite{Deelman2005}}       & explicit  & explicit  & configuration & script  & GUI & no & yes & multi-core/task & no  \\ \hline
\bf{Queue\cite{queue}}       & explicit  & explicit  & class-based & script   & CLI & no & partial & multi-core& no \\ \hline
\bf{Ruffus\cite{goodstadt2010ruffus}}       & explicit  & explicit  & convention & script & CLI & yes & no & multi-core/task;data partition & no  \\ \hline
\bf{Snakemake}\cite{koster2012snakemake}& implicit   & implicit    & convention & script & CLI  & yes & partial & multi-core & no \\ \hline
\bf{Swift\cite{tiberiu2007accelerating}}       & implicit   & implicit   & convention & script  & CLI & no & no & multi-core/task;data partition & no \\ \hline
\bf{Tavaxy\cite{Abouelhoda2012}}       & explicit  & explicit  & configuration  & GUI  & GUI & yes & partial & multi-core/task & no  \\ \hline
\bf{Taverna\cite{Oinn2004}}       & explicit  & implicit   & configuration & GUI & GUI & yes & yes & multi-core/task & no \\ \hline
\bf{Toil\cite{toil}}       & explicit   & implicit   & class-based & script & CLI & yes & no & multi-core/task & no \\ \hline
\end{tabular}
 \begin{tablenotes}
       \item[a] The tool dependency is {\em implicit} when the dependency among tools is automatically inferred from inputs and outputs, otherwise is explicit.
       \item[b] The execution order is {\em implicit} when is automatically inferred from a topological sort, otherwise is explicit.
       \item[c] The paradigm can be classified as {\em convention, configuration or class-based}. {\em Convention} defines frameworks that uses inline scripting code for task within a pipeline. When the description of tasks is configuration-based, the paradigm is classified as {\em configuration}. The paradigm is classified as {\em class-based} when the pipeline language is implemented in a class based language. The class-based frameworks are often closely bound to an existing code library.
       \item[d] The workbench that allows users to specify preconfigured modular tools together may have a graphical interface ({\em GUI}) or not ({\em script}). 
       \item[e] The pipeline execution can be done through the command-line ({\em CLI}) or through a graphical user interface ({\em GUI}). Notice that nearly all tools provide some kind of CLI, but it may not be the native way supported for the tool.
       \item[f] A nested pipeline is when it exists a pipeline that has a small subset of steps and is then connected to another pipeline.
       \item[g] A pipeline can be truly shared when a file describing it can be shared with another user that can execute it at a later time and without access to the same execution instance. A pipeline is said to partially shared when running depends on access to same execution instance or centralized system.
       \item[h] We split parallelization in three levels: multi-core execution of tool commands that support it, parallel execution of independent tasks of a pipeline, and data partitioning in order to process the partitions in parallel executions of the same pipeline.
       \item[i] With respect to containerization of tools, we distinguish the containerization of the framework as a whole from the containerization of tools independently, since the later one reflects a more interoperable framework. A yes here means that the system supports tool containerization.
 \end{tablenotes}
\end{threeparttable}
    }
\end{table}
\end{landscape}
\restoregeometry

We believe that a true framework should ensure the following characteristics.

{\em Reusability}. Pipeline specification should be independent of the execution environment in order to still be used whenever the execution environment differs. Moreover, pipelines should be reusable with different, but compatible, data files and algorithms.

{\em Reproducibility}. Reproducing experimental results is an essential facet of data analysis, providing the foundation for understanding, integrating, and extending results towards new discoveries.

{\em Transparency}. Users should be able to share and communicate experimental results and pipelines in a meaningful and independent way.

{\em Specification accessibility, without loosing flexibility}. The system should allow to define visually or programmatically each pipeline taking into account users heterogeneous backgrounds.

{\em Virtual environment setup accessibility}. The system should provide an easy way to set-up integration of local, grid, and cloud infrastructures as well as arbitrary compositions thereof.

{\em Anywhere execution}. One should be able to run a pipeline easily, using either local (even if only for small datasets) or remote computing facilities (such as dynamically scalable "pay-per-use" cloud computing infrastructures).

{\em Workload pattern awareness}. The execution engine should consider building blocks workload hints for optimizing the execution, making use of parallelization and adaptive scheduling.

{\em Interoperability}. The system should provide interfaces to support interoperability between different systems and data sources.

Therefore, working toward above requirements, we propose the NGSPipes framework based on three principles.
The first principle is to completely avoid servers and services configuration.
The second one is to automatically get and configure only required tools.
The third is to precisely describe pipelines.

The NGSPipes architecture and implementation relies on: 
a flow-based executable language for the specification of pipelines,
repositories for tools,
a virtual environment assembler,
and a standalone execution engine.
We developed also a user-friendly editor prototype for specifying pipelines as a proof of concept.

These components allow us to have an ubiquitous open system meeting most requirements above.
Namely, one of our main contributions is a pipeline specification language, with a clear separation between the language and the execution engine.
This language is suitable for end users with or without programming expertise, and without compromising the expressive power for describing pipelines (or more generally data flow  processing within a  directed acyclic graph model).
Moreover, by being system independent, we believe that the proposed language will allow pipelines to be transparently exported and reused within different systems in use.

The proposed framework aims also for the decoupling of concrete data and tools from workflows/pipelines specification.
This is particularly important if we take into account data privacy and tools licensing, essential issues for the scientific and industry communities.
 
The architecture of the proposed framework was designed to support the execution of pipelines without users need to configure, install and manage tools, servers and complex workflow management systems.
Moreover, given a pipeline to execute (described through the specification language), all the execution environment is automatically setup and the pipeline is executed.
Our current implementation does not take into consideration workload patterns and, hence, it does not use yet adaptive scheduling or automatic parallelization.

The remaining paper is organized in three main parts: framework architecture and implementation description, a case study, and discussion.
The framework and related prototypes are open source and readily available online.
%
%
%
\section*{Implementation}
\begin{figure}[!t]
\centerline{\includegraphics[width=\textwidth]{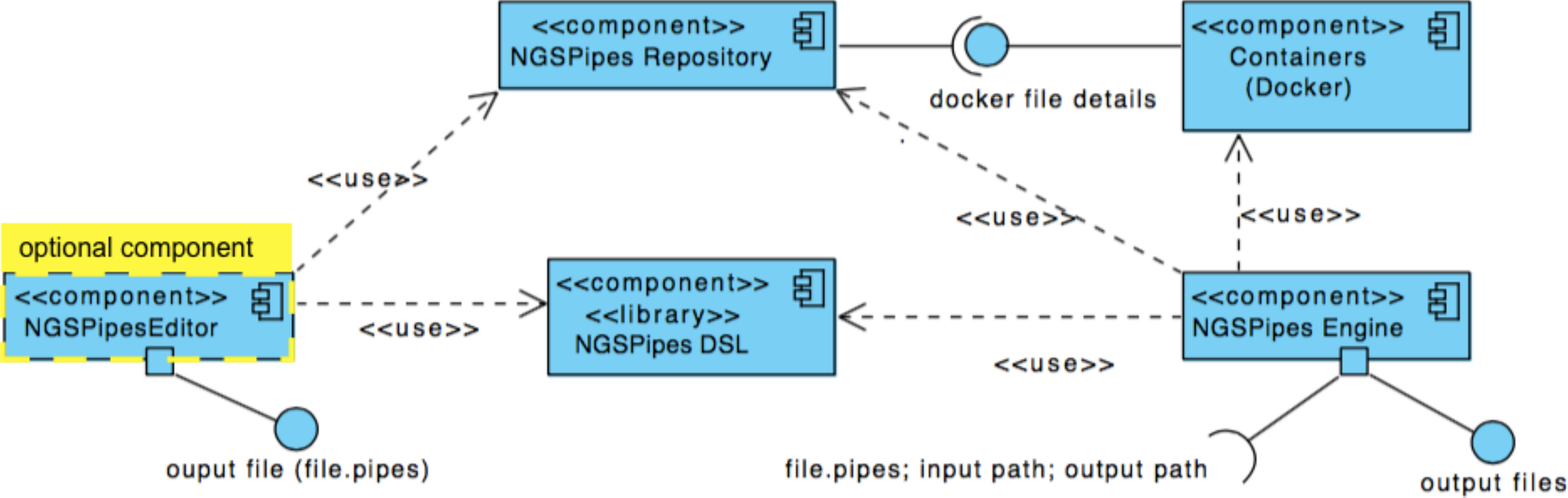}}
\caption{Component diagram that describes the overall architecture of the
NGSPipes framework.}
\label{fig:01}
\end{figure}

Let us introduce the {\em NGSPipes} framework. 
As shown in Figure \ref{fig:01}, the framework architecture comprises three main components:
\begin{itemize}
\item {\em The specification language}, a domain specific language (DSL) suitable for describing pipelines. 
\item {\em Repositories of tools} that contain the description of each tool available for integrating within pipelines.
    We note that new tools can be easily added and new repositories can be made available independently.
\item {\em The execution engine} which given a pipeline to execute (described using above language), automatically sets up all the execution environment and executes the pipeline.
\end{itemize}
All these components are independent, easily extensible and reusable, allowing a seamless integration of new tools for data analysis and processing.
Note that decoupling pipeline and tools specifications from data sources and real tools leads to a more flexible framework.
Moreover the use of repositories with version control allow us to attain both pipeline and tool description versioning.
We can for instance make available and run the pipeline using different, but compatible, versions of the same tool.
As we discuss later, we would only need to change the repository of tools being referenced by the pipeline specification.

The architecture of the execution engine provides also a self-contained and decentralized framework.
This aspect together with the leveraging of public repositories allow the sharing and use of pipelines in the medium/long-term.

\begin{figure}[!t]
\centerline{\includegraphics[width=\textwidth]{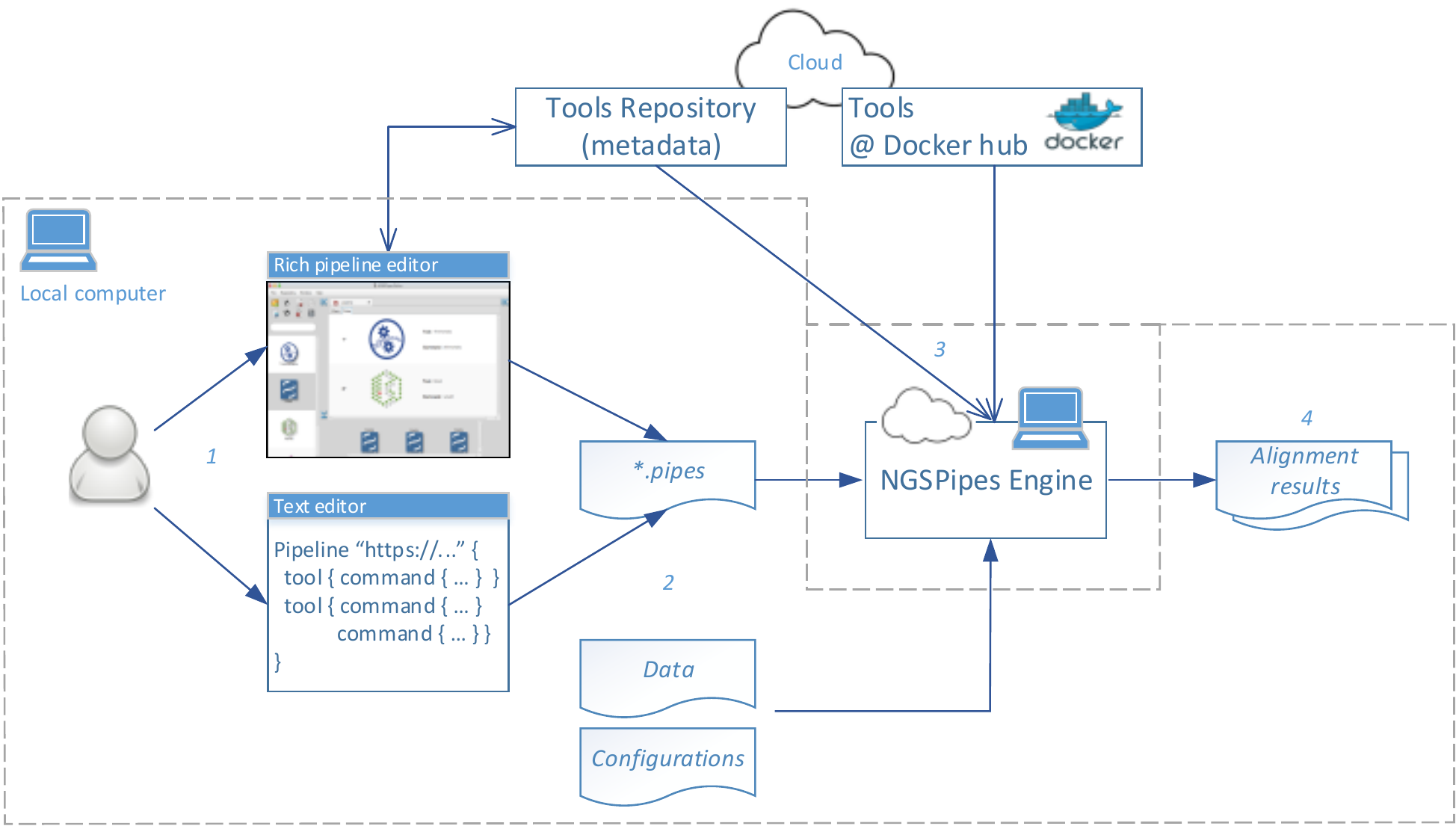}}
\caption{NGSPipes interaction diagram. In step 1 the user defines a pipeline ether using a text editor or a rich editor. In step 2 the user produces the \texttt{.pipes} file, a step only necessary when defining it with the rich pipeline editor. 
In step 3 the user executes the pipeline using the NGSPipes engine. In step 4 the user get the results after execution has finished.}
\label{archflow}
\end{figure}

A diagram with the interaction flow of the NGSPipes framework can be found in Figure~\ref{archflow}.

\subsection*{Tools Repository}\label{sec21}
Each {\em repository of tools} contains all the information related to a set of available tools for constructing pipelines (see Figure \ref{fig:01}).
Such information includes details as what is necessary to install and/or execute a given tool, and where we can fetch it.
Thus, the pipeline definition does not need to include these details and, on the other hand, we are able to automatically assemble the execution environment. 

For each available {\em tool}, the repository should include a {\em tool descriptor} and at least a {\em tool configurator}.
The {\em tool descriptor} is the entity responsible for supplying all the information on how to run a given tool, such as available commands and arguments, processor options and memory requirements.
This information should be described according to the specification documentation~\cite{ngspipestools} and discussed below.
Let us take as an example the tool Velvet~\cite{Zerbino2008}.
It includes the commands \verb+velvetg+ and \verb+velveth+, and a fragment of its descriptor is shown in Figure~\ref{fig:03}.
\begin{figure}[!t]
\begin{lstlisting}[language=metadata]
 "name" : "Velvet",
 "version" : "0.7.01",
 "setup" : [ "make" ],
 "requiredMemory": 12288,
  "commands" :[
       { "name" : "velveth",
         "command" : "velveth",
         "priority" : 2,
         "arguments" : [
           { "name" : "output_directory",
             "outputType" : "outputDir",
             "isRequired" : "true",
             ...
           },
	       ...
	   }
	   ... 			
\end{lstlisting}
\caption{Partial descriptor for Velvet tool.}
\label{fig:03}
\end{figure}
A descriptor must include at least:
the tool {\em name},
the tool {\em version},
memory requirements ({\em requiredMemory}),
{\em setup} scripts to be executed on execution environment setup, and available {\em commands}.
Each command is described following a similar approach:
the command {\em name},
the (real) {\em command} to be executed,
the {\em priority} of the command,
{\em arguments} and {\em outputs} generated,
and the argument composer ({\em argumentComposer}) for specifying how to link arguments to values.
 

A {\em tool configurator} includes the information needed to define the execution context for a given tool:
the {\em name} of the file where the execution context is defined,
the name of the execution context (the {\em builder}),
the {\em setup} scripts that must be executed for assembling the execution context,
and the {\em uri} that identifies and allows to fetch the tool.
Figure \ref{fig:05} shows an example where the tool is provided by a docker image and, thus, it is necessary to install docker in the execution context~\cite{Docker}.
\begin{figure}[!t]
\begin{lstlisting}[language=metadata]
  { "name" : "DockerConfig",
    "builder" : "Docker",
    "uri" : "ngspipes/velvet0.7",
    "setup" : [
		"wget -qO- https://get.docker.com/ | sh"
    ]
  }
\end{lstlisting}
\caption{An example of a configurator for the Velvet tool.}
\label{fig:05}
\end{figure}

The implementation of the repository component provides an interface in order to be extensible.
New repositories can then be made available and should include: 
a list of {\em tool descriptors},
a list of {\em configurators} for each tool,
the configurators for each tool named accordingly,
and a logo for each tool.
We provide an implementation of the repository component also as a support library.
This support library includes an implementation for a remote repository on Github and for local repositories.
An example of repository is available and can be found in tools documentation.
Note that if users define their own remote Github repository using the same schema, then they do not need to extend the support library.
This is only required for new types of repository. More information about this schema can be found in tools documentation.
 
\subsection*{Specification Language}\label{sectionlanguage}\label{sec22}
The {\em specification language} is a DSL for describing pipelines. 
It contains primitive building blocks with the enough expressiveness to define data processing pipelines, namely when data processing can be modelled as a directed acyclic graph. 
Figure~\ref{fig:06} depicts partially the syntax of this language, given by a grammar. 
The full syntax of the language can be found in DSL documentation~\cite{ngspipesdsl}, using an EBNF notation alike.
\begin{figure}[!t]
{\em pipeline} ::= {\bf Pipeline} {\em repositoryType} {\em repositoryLocation} {\bf $\{$} ({ \em tool}) + {\bf $\}$};\\
{\em tool} ::= {\bf tool} {\em toolName} {\em configurationName} {\bf $\{$} ({\em command})+ {\bf $\}$};\\
{\em command} ::= {\bf command} {\em commandName} {\bf $\{$} ({\em argument} | {\em chain}) + {\bf $\}$};\\
{\em argument} ::= {\bf argument} {\em argumentName} argumentValue;\\
{\em chain} ::= {\bf chain} {\em argumentName} (({\em toolName})? {\em commandName})? {\em outputName};
\caption{Partial grammar for the specification language using EBNF notation.}
\label{fig:06}
\end{figure}
The primitives of the language are {\em Pipeline}, {\em tool}, {\em command}, {\em argument} and {\em chain}. 
Since a {\em Pipeline} implies the execution of one or more tools, its specification must reference the tools repository that is being used.
 
The reference to the repository found in the specification of a pipeline must identify not only where to find the repository, but also the type of repository: local or remote, like Github.
As shown in Figures~\ref{fig:06} and \ref{fig:07}, the first line in a pipeline specification states the type of repository (\verb+Github+) and where the repository can be found.
\begin{figure}[!p]
\begin{lstlisting}[language=ngspipes]
Pipeline "Github" "https://github.com/ngspipes/Repository" {
  tool "Trimmomatic" "DockerConfig" {
    command "trimmomatic" {
      argument "mode" "SE"
      argument "quality" "-phred33"
      argument "inputFile" "ERR406040.fastq"
      argument "outputFile" "ERR406040.filtered.fastq"
      argument "fastaWithAdaptersEtc" "adapters/TruSeq3-SE.fa"
      argument "seed mismatches" "2"
      argument "palindrome clip threshold" "30"
      argument "simple clip threshold" "10"
      argument "windowSize" "4"
      argument "requiredQuality" "15"
      argument "leading quality" "3"
      argument "trailing quality" "3"
      argument "minlen length" "36"
    }
  }
  tool "Velvet" "DockerConfig" {
    command "velveth" {
      argument "output_directory" "velvetdir"
      argument "hash_length" "21"
      argument "file_format" "-fastq"
      chain "filename" "outputFile"
    }
    command "velvetg" {
      argument "output_directory" "velvetdir"
      argument "-cov_cutoff" "5"
    }
  }
  tool "Blast" "DockerConfig" {
    command "makeblastdb" {
      argument "-dbtype" "prot"
      argument "-out" "allrefs"
      argument "-title" "allrefs"
      argument "-in" "allrefs.fna.pro"
    }
    command "blastx" {
      chain "-db" "-out"
      chain "-query" "Velvet" "velvetg" "contigs_fa"
      argument "-out" "blast.out"
    }
  }
}
\end{lstlisting}
\caption{Example of a pipeline specification.
See the section on the case study for details concerning this pipeline.}
\label{fig:07}
\end{figure}
Each tool used in a pipeline is then specified by providing its name, its configurator and the list of tool commands that will be executed within this pipeline.
For instance, in the pipeline of Figure~\ref{fig:07} the second tool is the Velvet tool, and DockerConfig is the chosen configurator.
This information together with the repository information specifies the environment for executing Velvet commands.
Note  that the same command for a given tool may be executed several times and with different parameters, being listed more than once.
Note also that the commands within different tools may be interleaved.

As mentioned before, each command in the pipeline appears in the context of a tool.
For executing each command, it is necessary to identify its name, which is unique in the tool context, and to set the arguments for required parameters.
For instance, in the pipeline of Figure~\ref{fig:07}, the argument \verb+file_format+ for command \verb+velveth+ has as argument ``-fastq'', {\em i.e.}, the input file for this command must be in FASTQ format.

The specification language also includes the {\em chain} primitive for linking outputs into inputs.
With this primitive we can define as an argument of a command an output file of other command.
This primitive is used to specify execution flows.
The output from each command may be files named internally by the command or named through command arguments.
In both situations it is common that other commands use these output files for keep processing the pipeline.
For instance, in command \verb+blastx+, the argument \verb+-query+ receives as value the file ``contigs$\_$fa'', which is an output of the command \verb+velvetg+ of tool \verb+velvet+.
The {\em chain} primitive has a simplified version, which can be used when the output is from the previous command in the pipeline specification.
In this case, we only specify the name of the output file to chain with the given argument.
As an example, we can see in Figure~\ref{fig:07} the argument \verb+filename+ of \verb+velveth+ command chained with the output file, named as ``outputFile'', of command \verb+trimmomatic+.

\begin{figure}[!t]
\centerline{\includegraphics[scale=0.9]{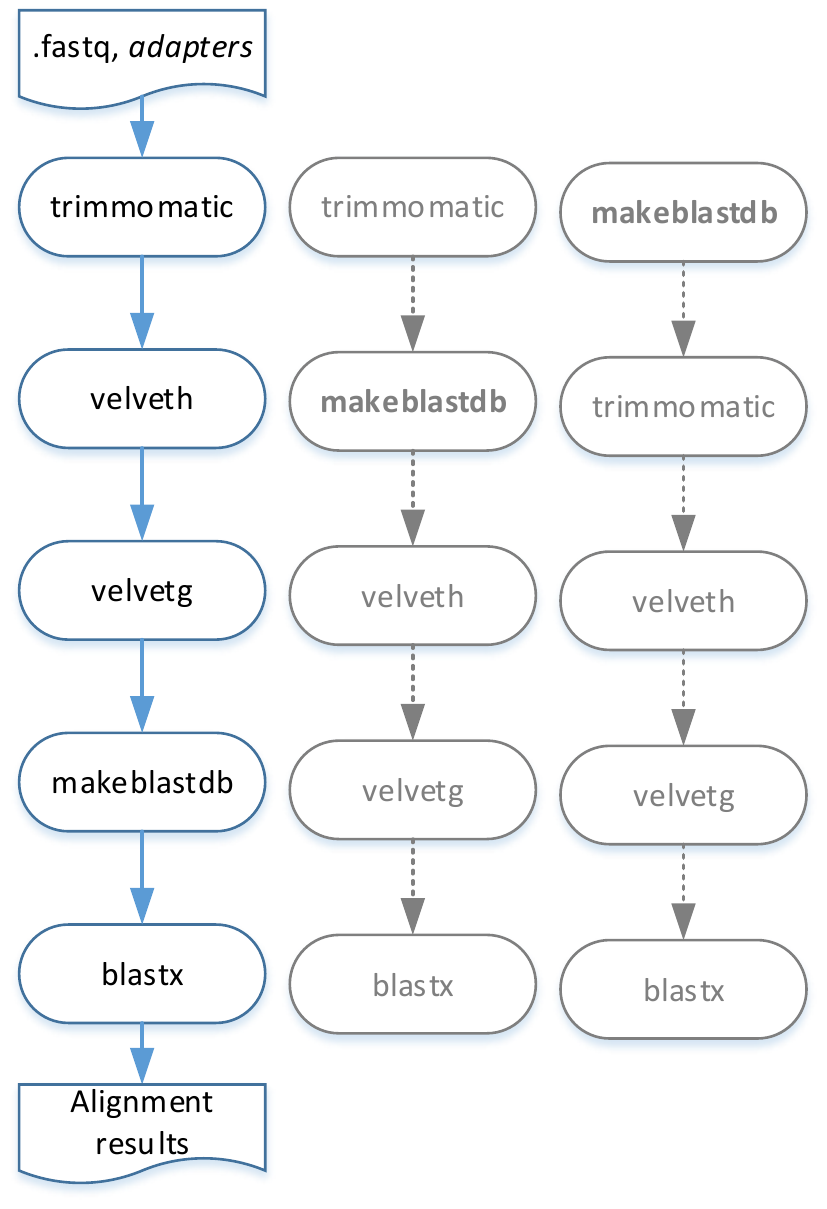}}
\caption{An execution order chosen among several possibilities inferred within the language library ahead of execution.}
\label{fig:ot}
\end{figure}

A library was developed to support the execution of this new language, through which pipeline specifications are internally mapped to Java code.
Given a pipeline and the underlying tool execution dependency graph, the library also infers one of possible execution orders by computing a topological order of that graph, as depicted in Figure~\ref{fig:ot} for the running example.
This is an automatic step.
%
%
The engine, described in the next section, uses this library to execute the pipeline.

\subsection*{Engine}\label{sec23}
The {\em engine} is responsible for:
the analysis of the pipeline description and transformation to an executable format,
the setup of tools used in the pipeline description,
and for the execution of the tools in an isolated context.

The pipeline description is transformed to an executable format.
Because the pipeline can be specified outside the editor, language consistency checks must be applied.
For the setup of the execution environment, the engine relies on information collected from the repository of tools referred in the pipeline specification. 
The orchestration and planning of execution is delegated to the language library. 
It checks the correct execution of the pipeline steps and outputs the relevant information to the user.

\begin{figure}[!t]
\centerline{\includegraphics[width=\textwidth]{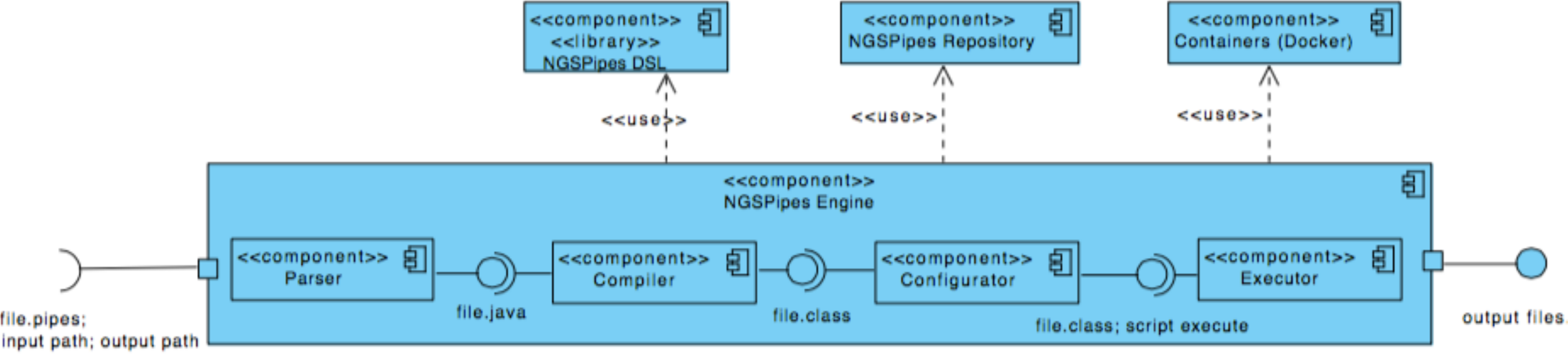}}
\caption{Component diagram of the execution engine.}
\label{fig:09}
\end{figure}

Figure~\ref{fig:09} shows the main components of the {\em engine} and their interaction.
The engine receives the pipeline description, the input path and the output path.
Internally, the {\em engine} is divided in four components.
The parser transforms a pipeline (described in the language presented above) to a representation in the Java language~\cite{javabook}.
Any grammatical errors are detected in this phase.
The second component is the {\em compiler}, which will produce an executable pipeline with the correct invocation sequence. 
Note that no tools are embedded in this executable pipeline.
They will be dynamically downloaded and executed only by the {\em executor}.
The third component is the {\em configurator}, which is responsible for the configuration of the {\em executor} and for booting the pipeline execution phase.

The configuration data consists of the computational resources that will be available during execution (i.e. amount of memory and number of CPU cores) as well as input and output paths.
Part of the configuration data is obtained automatically by looking at the executable pipeline and the {\em repository of tools}.
By looking at the pipeline, the {\em configurator} determines which tools where used and, by looking at the {\em repository}, it determines the memory required to run each tool.
The amount of memory set by the {\em configurator} will be the highest value among all the included tools.

The fourth component of the {\em engine} is the {\em executor} and it relies on two layers of virtualization.
The first layer is a system-level type of virtual machine (VM).
The current framework implementation relies on a widely used hypervisor to run this VM -- the VirtualBox system, as described in engine documentation~\cite{ngspipesengine}.
This allows the engine to be installed on any type of main stream operating system (e.g. macOS, Windows, GNU/Linux).
Inside the virtual machine, a Linux-based operating system is ready to be executed.
On top of this, the {\em engine} uses a lightweight virtualization technology to ensure proper installation, keep up-to-date, and run each command of the pipeline. 
Currently, the framework uses Docker containers technology~\cite{Docker}.
Other solutions can be integrated in the future because both the pipeline language and repositories of tools are not compromised with this technology. 

Before the actual steps of the pipeline are executed, the {\em engine} ensures that the VM is ready to use the container technology by checking if necessary packages are available.
Once completed, the pipeline is executed, downloading and running the correct tool/command.
The download part is done only in the first execution.
After that, tools remain installed and ready for later executions.
Because each command is executed in a separated container, the {\em executor} must ensure that the input files, located in the user environment, are made available to each tool.
 
\begin{figure}[!t]
	\centering
	\begin{minipage}[b]{0.5\linewidth}
		\centering
		\includegraphics[width=\textwidth]{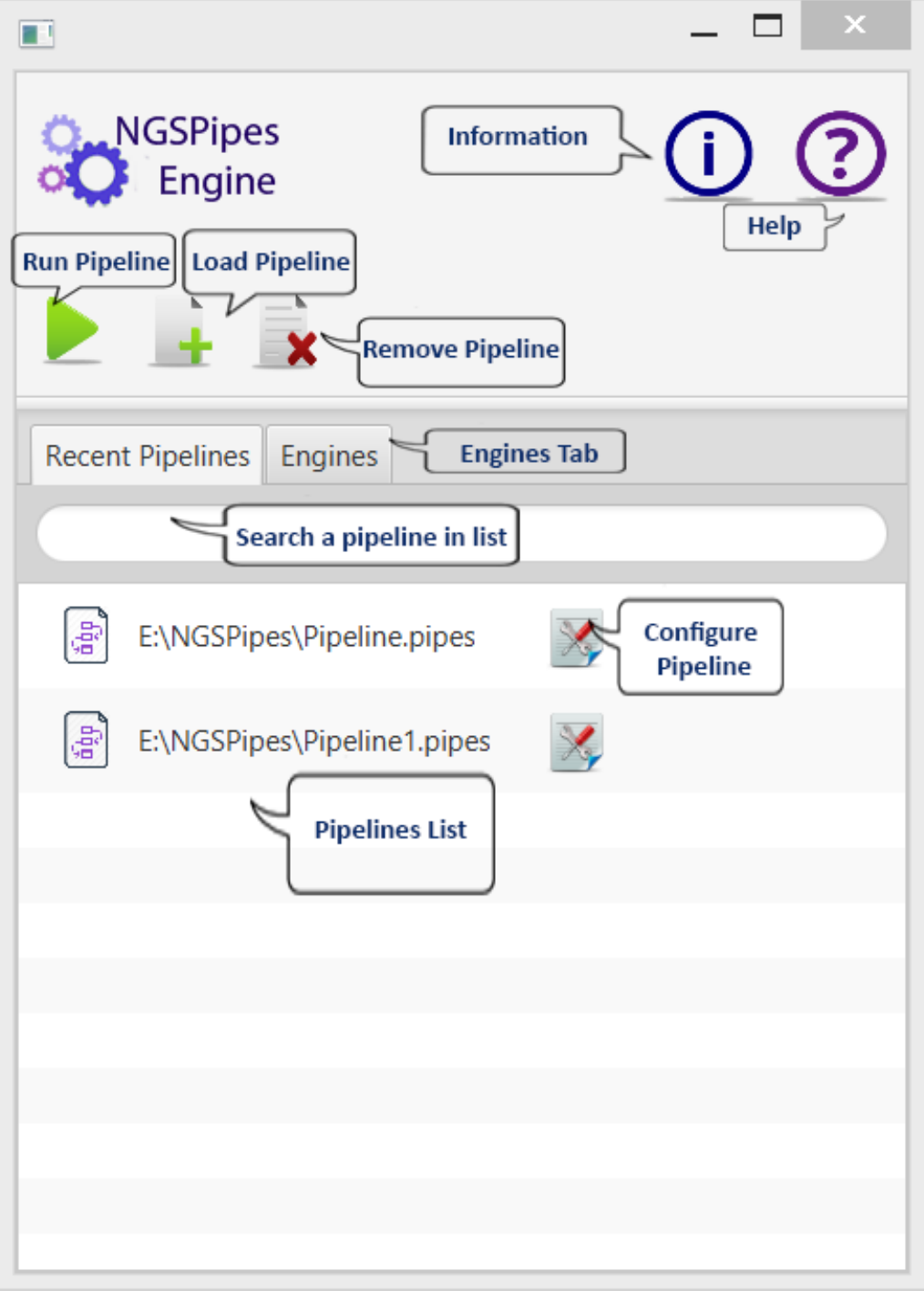}
		\\	
		\textsf{a.}
	\end{minipage}	
	\hspace{0.5cm}
	\begin{minipage}[b]{0.4\linewidth}
		\centering
		\includegraphics[width=\textwidth]{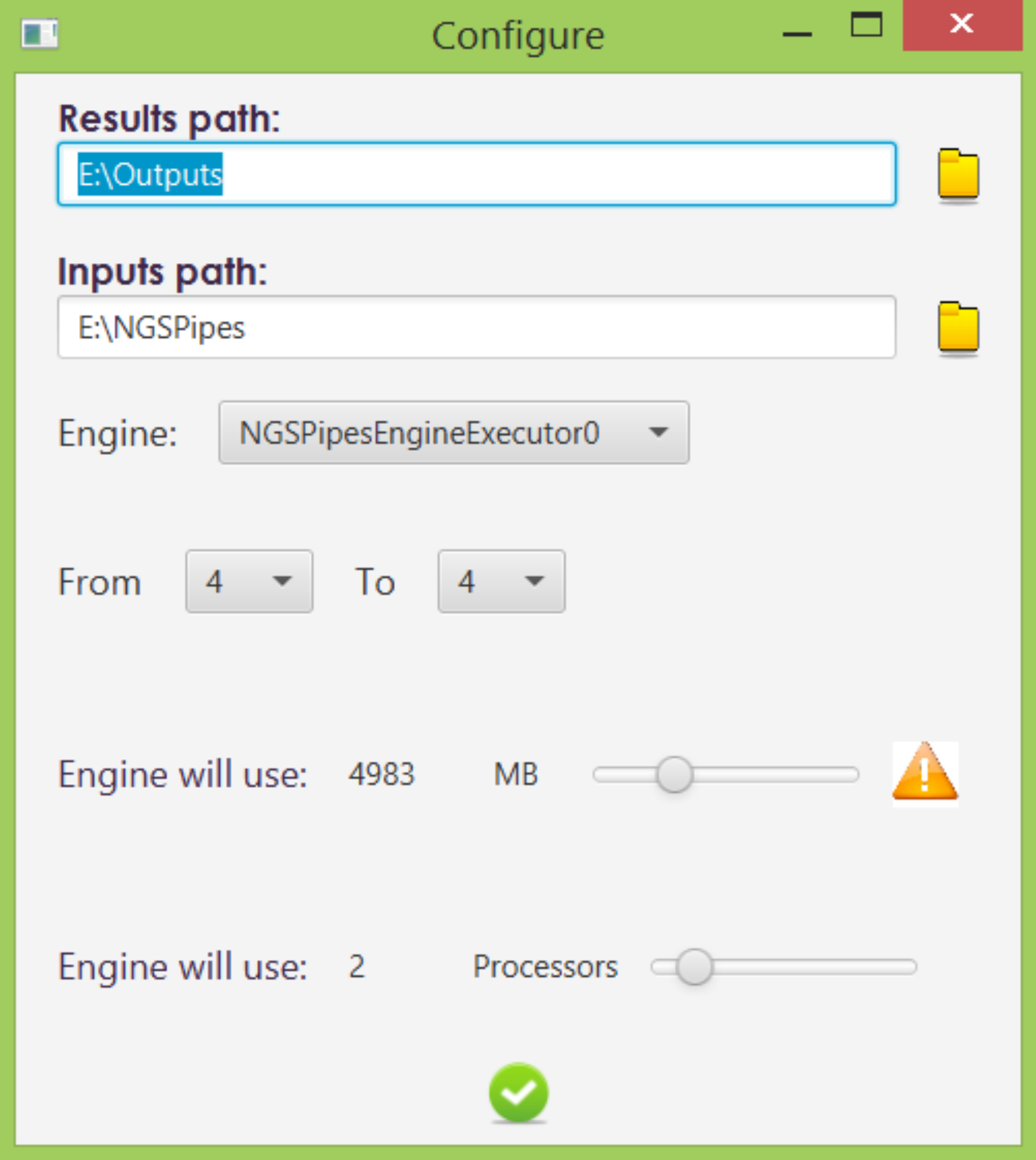}
		\\
		\textsf{b.}
	\end{minipage}
	\caption{NGSPipes engine. (a) Main window. (b) Configuration window.}
	\label{fig:engine-UI}
\end{figure}

The {\em engine} is available in two versions: a command line application and a graphical user interface (GUI) application. Both versions are functionally equivalent and are packed as a regular Java application.
Figure ~\ref{fig:engine-UI}.a) shows the operations available -- run, load and remove a given pipeline.
The execution of a given pipeline can be parametrized, including the input and output directories as well as hardware resources made available to the pipeline execution.
The window to do that is presented in Figure~\ref{fig:engine-UI}.b).
The amount of memory can be limited, which overrides the value determined automatically by the {\em Configurator}.
However, doing so can result in an execution error if insufficiency memory is specified for the data to be processed.  To change the amount of memory is essential to know how much data is going to be processed. For instance, although it is recommended having 12 GBytes of physical memory for using the Velvet tool, in our case study we will only need 4 GBytes of memory
since for bacterial strain sequencing we can use less memory in general.
Multiple pipelines can be executed simultaneously.
This is done in a transparent way to the user, and is only limited to the physical amount of resources (i.e. memory and CPU) available at the computer where the pipeline is running.
When starting a pipeline, the engine determines if other execution is already ongoing.
If so, a new {\em executor} is created and the pipeline starts executing in this new instance.
This feature can be particularly useful when running on a multi-core multi-gigabytes machine, either a physical server in a private datacenter, or a virtual machine located in a public cloud provider.
In this scenario, the command line version is the recommend way to use the system.

\subsection*{Language editor prototype}
An editor prototype was also implemented for producing and editing pipeline specifications.
The editor is focused on simplicity allowing users to describe and define graphically their data processing steps as building blocks.
Each block, depending of their corresponding type (e.g., the algorithm to be executed), may be parameterized through the visual interface.
The editor prototype was developed using the JavaFX library.

When we use the editor prototype instead of a text editor, all repository information is automatically imported to the editor when the user specify it.

For each new pipeline, the editor generates two files.
A file with extension \verb+.pipes+ which has the specification of the pipeline in above specification language, and a XML file which includes all the information that is only needed for the editor, such as the visual location of the tool boxes on the editor.
We note that the file \verb+.pipes+ is only generated/updated on demand.
In particular, when the pipeline definition is completed, the user must explicitly end it using the ``generate'' instruction.
This is important for defining the input files directory and for assuring that all files are copied to that directory.
As mentioned before, it is possible to use a text editor instead of the this editor.
In that case the user must define the pipeline, using pipeline primitives and must create an input directory, where all the input files must be, and an output directory, where all the pipeline output will be redirected.
The paths to these directories are not defined in the pipeline, they are given instead as parameters to the engine component.

\section*{Case study}
\label{section:case_study}
We consider a standard pipeline used on epidemiological surveillance using NGS data.
The aim is to characterize bacterial strains through allelic profiles~\cite{Maiden2013}.
This case study allows us to illustrate our approach and to evaluate performance overheads.
Other examples can be found in the online documentation.

When sequencing a bacterial strain by paired end methods with desired depth of coverage of 100x (in average each position in the genome will be covered by 100 reads), the output from the sequencer will be two FASTQ files containing the reads.
Each read typically will have 90-250 nucleotides length, using Illumina technology.
The first data processing step is to trim the reads for removing the adapters used in the sequencing process and any tags used to identify the experiment in a run.

From clean reads two approaches can be followed: {\em de novo} assembly or mapping to a reference genome.

In {\em de novo} assembly, software such as Velvet~\cite{Zerbino2008} or SPAdes~\cite{Bankevich2012} is used to obtain a draft genome composed of contigs, longer DNA sequences resulting from assembling multiple reads.
Annotation software such as Prokka~\cite{Seemann2014} can then take contigs as input and determine the gene content and annotate it against multiple databases.
Alternatively, the draft genome can be compared to databases of gene alleles for multiple loci using BLAST~\cite{Altschul1990}.
Given BLAST results we can create an allelic profile characterizing the strain~\cite{Maiden2013}.

In mapping approaches, a reference genome is chosen and the reads are directly mapped against it using read mapping software such as BWA~\cite{Li2010} or Bowtie2~\cite{Langmead2012}.
The output is a file containing the relative position of each read in the reference genome.
That file is then processed to determine the positions that have single nucleotide polymorphisms (SNPs) when compared to the reference genome~\cite{Li2009}.
The resulting SNPs are then analyzed to determine if they might be the result of recombination events~\cite{Croucher2014}, and filtered out if they are to be used in phylogenetic analysis.
Several allelic or SNP profiles for different strains resulting from both approaches can then be compared to determine their phylogenetic relationships using different methods~\cite{Huson1998,Francisco2009}.

\begin{figure}[!t]
	\centering
	\includegraphics[width=\textwidth]{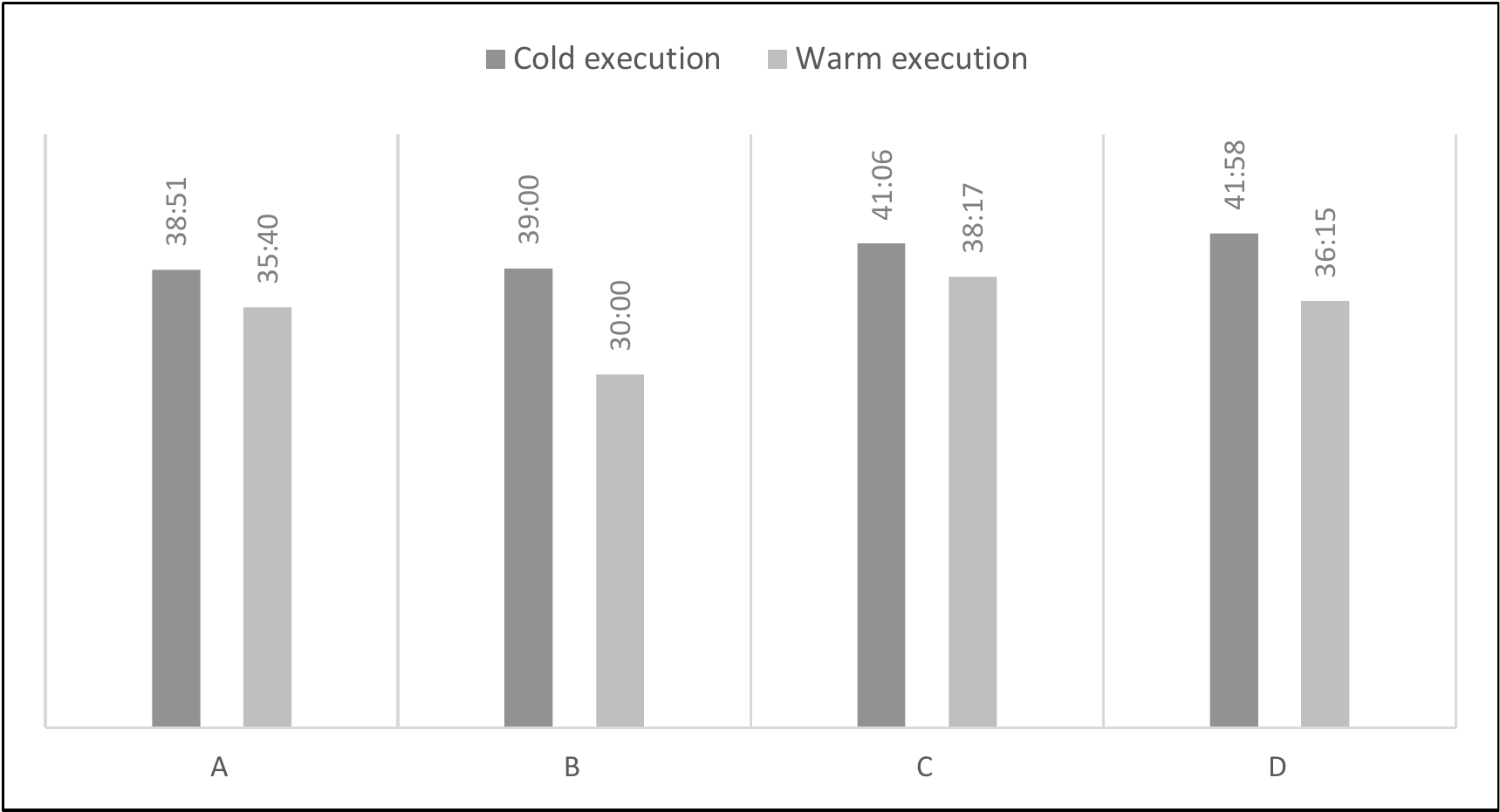}
	\caption{Performance of cold and warm execution}
	\label{fig:performance}	
\end{figure}

The pipeline in Figure~\ref{fig:07} follows the {\em de novo} assembly approach and relies on BLAST for comparing the draft genome to a database of gene alleles.
We relied on data from NCBI Sequence Read Archive for testing and evaluating the framework, namely data on {\em Streptococcus pneumoniae}. 
See the use case documented within engine documentation~\cite{ngspipesengine} for more details.
Figure \ref{fig:performance} shows execution times when running this pipeline. The systems used to run the pipeline differ in hardware and operating system, as presented in Table~\ref{tab:systems}.
\begin{table}[!t]
\centering
\caption{Operating system and hardware of four different experimental setups.}
\begin{tabular}{|l|l|l|l|l|}
	\hline
	System & OS & CPU & RAM & Disk type \\ 
	\hline
	A & Windows 10 & Intel(TM) i5 2.5 GHz & 8 GB & SSD \\
	B & Windows 10 & Intel(TM) i7 2.5 Ghz & 16 GB & HDD \\
	C & OS X Yosemite & Intel(TM) i5 1.8Ghz &  8 GB & SSD \\                 
	D & Slackware 14.0 &  AMD Opteron(TM) 2.3GHz & 256 GB & HDD    \\                       
	\hline
\end{tabular}
\label{tab:systems}
\end{table}
To evaluate the engine performance we execute it assigning 2 cores (parameter \verb|-cpus 2|) and 4 GBytes of RAM (parameter \verb|-mem 4|).  The size of the FASTQ input file is 814 MBytes. The pipeline (see Figure~\ref{fig:07}) and initial data was used a first time, which we call \textit{cold execution}. During this run, the engine automatically installs the necessary tools and keeps them installed for the second and following executions, which we call \textit{warm execution}. Figure~\ref{fig:performance} depicts the results for these two scenarios and the four systems described in Table \ref{tab:systems}. Depending on the system, the pipeline takes between 38 and 42 minutes to execute. We also note that keeping the tools installed for new executions is a good option since the speedup of a warm execution varies between 7\% (system C) and 23\% (system B).
It is outside the scope of this paper to further investigate the distinct performances that can be observed across different operating systems. It is however relevant to understand if the organization of the engine introduces a significant overhead to the native execution of the tools. Besides the hypervisor -- VirtualBox -- other sources of overhead include the thin layer of virtualization introduced by Docker. This technology uses linux containers to run each image, and studies have shown that linux containers have a negligible overhead~\cite{Xavier2013}. So, we expect that VirtualBox is the only point of some performance loss when compared to direct execution of each tool. Nevertheless, this overhead should be small~\cite{Langer2011} while keeping the advantage of having an highly portable and up-to-date NGS pipeline execution engine.

\section*{Discussion}
Although much NGS data have been published and shared in recent years, we cannot yet talk about open science.
Even if tools are available, analyses pipelines are often not detailed and clearly defined.
This includes tools parametrization. Or pipelines cannot be shared and archived for future reference.
Hence, it is almost impossible to reproduce and validate published results, or to use exactly the same approach with different data.

This is a well known problem and, as mentioned in introduction, many platforms and workflow engines have been proposed (see Table~\ref{tablecomparison}).
Still these platforms depend in general on installing and having available suitable servers with all possible tools configured.
See for instance Galaxy~\cite{Giardine2005}.
In general this is a problem for many researchers and professionals, even for those teams with professional IT support.
Given all software dependencies and versioning the task is far from trivial.
Even more recent engines and frameworks, such as Bpipe~\cite{Sadedin2012}, Snakemake~\cite{Koster2012}, or Nextflow~\cite{nextflow}, require either a unix-like environment or that all tools are locally available and configured.
On the other hand given the state of the art in what concerns computer systems and engineering, this should no longer be a problem.
NGSPipes framework addresses this issue by making use of such state of the art technologies, such as containerization and virtualization.

NGSPipes framework makes use of resources and environment isolation for making tools available, avoiding servers and services manual configuration.
This approach is well known in IT industry and is already being adopted for life sciences~\cite{BioShaDock}.
Such ecosystem is of crucial importance for NGSPipes framework and is being also adopted by other platforms, e.g., Galaxy~\cite{Giardine2005} and Nextflow~\cite{nextflow}.
This is an important step since we can share resources among many different systems and platforms.
Note in particular that NGSPipes framework is agnostic with respect to the job performed by each tool.
A tool can invoke remote Web services, fetch remote data, or even make use of cloud computing resources either directly or through other workflow systems.

Still tools alone are not of much use, analyses pipelines must be made available, precisely defined, and platform independent.
In this paper we propose a simplified specification language and support library for this purpose, based on a clear and straightforward syntax.
Note that both language and library are completely independent from the execution engine.
In particular it can be reused by any other platform.
More general specification languages exist for specifying pipelines and workflows, being BPMN 2.0 the most well known, which specification
and documentation can be found at \url{http://www.bpmn.org/}.
BPMN 2.0 allows however to model more complex processes, such as long running business transactions, and it possibly introduces another layer of complexity for life sciences practitioners.
Still we believe that the language proposed in this paper would benefit from being mapped and aligned with a subset of BPMN 2.0.

%
The main aim and novelty of NGSPipes framework is to provide a decoupled architecture for automatically setup and run pipelines without requiring users to deal with low level details of computer systems.
A user can access to a pipeline made available with a given study and just run it on the same data used in that study.
The pipeline can even be prepared to fetch original data from archival repositories such \url{http://zenodo.org}.
And the pipeline itself, as well as tool descriptors, can be stored in public repositories.
If we use repositories with versions like well known git services (e.g. \url{http://github.com}, \url{http://gitlab.com} or \url{http://bitbucket.org}) or archival repositories, then both pipelines and tool descriptors get versions automatically.
New or updated pipelines can be build from existing repositories, and new or updated tools can be made available by just releasing a new descriptor.
Tools and their different versions should be also publicly available, which is the case with most open source tools.
The fundamental concept is not to change existing pipelines, but to revise and extend them, leading to new versions.
But such approach can only be achieved through a decentralized framework like the one proposed in this paper.
As we discussed above, most tools available rely on centralized instances or services that must be maintained, what can be problematic in the medium/long-term.
This is of particular relevance even if we use frameworks like NGSPipes.
A pipeline can use tools that rely on Web services, a dependency that may lead to a broken pipeline once the Web service is no longer available.
We may also loose the pipeline versioning as the Web service may not be available in different versions.
This is however a user design decision that we cannot avoid.
Nevertheless a self-contained framework, that relies on widespread technologies and on widely used repositories, is an important piece to allow pipelines to be archived and used.

Although in present version NGSPipes framework can be easily used and integrated in cloud services since it relies on common cloud technologies, some issues remain.
As raised in introduction, deployment on cloud should be transparent, in fact we would say that executing a pipeline should be as easy as downloading a file from Web.
There is however work to be done as cloud technologies mature and become commodity.
Task parallelization and distribution is another issue.
The heterogeneous nature of NGS data and analyses jobs, relying on different tools, lead to rather different computational workloads.
In this context both task scheduling and resources provision planning should be aware of workload patterns.


\section*{Acknowledgements}
  This work has been partially supported by national funds through FCT -- Funda\c{c}\~{a}o para a Ci\^{e}ncia e Tecnologia, under
INESC-ID strategic project (UID/CEC/50021/2013).

\bibliographystyle{bmc-mathphys} 
\bibliography{article}      

\end{document}